\def\ang{\AA}

\def\gapprox{\lower.4ex\hbox{$\;\buildrel >\over{\scriptstyle\sim}\;$}}
\def\lapprox{\lower.4ex\hbox{$\;\buildrel <\over{\scriptstyle\sim}\;$}}

\def\refer#1   {\par\noindent\hangindent1cm {#1}}

 \documentstyle[aaspp4]{article}        
 \input epsf
\slugcomment{}
\lefthead{}
\righthead{}

\begin{document}
{\sl The Astrophysical Journal, Vol. {\bf 598}, Dec 1, 2003, issue (in press).}

\title{         Observational Tests of Damping by Resonant Absorption in Coronal Loop Oscillations	}
		
\author{        Markus J. Aschwanden and Richard W. Nightingale}
\affil{         Lockheed Martin Advanced Technology Center,
                Solar \& Astrophysics Laboratory,
                Dept. L9-41, Bldg.252,
                3251 Hanover St.,
                Palo Alto, CA 94304, USA;
                e-mail: aschwanden@lmsal.com}
\and
\author{	Jesse Andries, Marcel Goossens, and Tom Van Doorsselaere }
\affil{		Centre for Plasma Astrophysics, K.U.Leuven,
		Celestijnenlaan 200B, 
		3001 Leuven, Belgium;
		e-mail: jesse.andries@wis.kuleuven.ac.be,
		marcel.goossens@wis.kuleuven.ac.be,
		tomvd@wis.kuleuven.ac.be}

\begin{abstract}
One of the proposed damping mechanisms of coronal (transverse) loop oscillations in the kink-mode
is resonant absorption as a result of the Alfv\'en speed variation at the outer boundary 
of coronal loops. Analytical expressions for the period and damping time exist for loop 
models with thin non-uniform boundaries. They predict a linear dependency of the ratio 
of the damping time to the period on the thickness of the non-uniform boundary layer. 
Ruderman and Roberts used a sinusoidal variation of the density in the non-uniform 
boundary layer and obtained the corresponding analytical expression for the damping time. 
Here we measure the thickness of the non-uniform layer in oscillating loops for 11 events, 
by forward-fitting of the cross-sectional density profile $n_e(r)$ and line-of-sight
integration to the cross-sectional fluxes $F(r)$ observed with TRACE 171 \ang . 
This way we model the internal $n_i$ and external electron density $n_e$ of the 
coronal plasma in oscillating loops. This allows us to test the theoretically predicted 
damping rates for thin boundaries as function of the density ratio $\chi = n_e/n_i$. 
Since the observations show that the loops have non-uniform density profiles 
we also use numerical results for damping rates to determine the value of $\chi$ for the loops.
We find that the density ratio predicted by the damping time,
${\chi}_{LEDA}=0.53\pm0.12$, is a factor of $\approx 1.2-3.5$ higher than the density ratio 
estimated from the background fluxes, ${\chi}=0.30\pm 0.16$. The lower densities modeled 
from the background fluxes are likely to be a consequence of the neglected hotter plasma 
that is not detected with the TRACE 171 \ang\ filter. Taking these correction into account, 
resonant absorption  predicts damping times of kink-mode oscillations that are commensurable 
with the observed ones and provides a new diagnostic of the density contrast of oscillating loops.
\end{abstract}


\section{		INTRODUCTION 				}

Oscillations of coronal loops have now been detected virtually in all wavelengths (for a 
recent review see, e.g., Aschwanden 2003). Most of these oscillations have been
interpreted in terms of standing (eigen-modes) and propagating MHD waves (for a recent
theoretical review see, e.g., Roberts \& Nakariakov 2003). The MHD eigen-modes include
fast sausage and kink modes that produce transverse oscillations with Alfv\'enic speed, 
slow magneto-acoustic modes that produce longitudinal oscillations with sound speed,
as well as torsional modes that produce sheared azimuthal oscillations. Obviously,
observations of such oscillating systems provide direct measurements of Alfv\'en speeds 
and sound speeds, which in combination with electron density measurements can be used 
to infer the magnetic field in the corona, which is very difficult to determine by other
means. This important new diagnostic has been dubbed {\sl coronal seismology} (Roberts,
Edwin, \& Benz 1984; Roberts \& Nakariakov 2003). 

Most of the coronal loops that exhibit oscillations have been found to be strongly
damped, typically having an exponential damping time $t_D$ of a few oscillation periods $P$
(Nakariakov et al. 1999; Schrijver et al. 2002; Aschwanden et al.
2002). Theoretical models of damping mechanisms include: (1) non-ideal effects such
as viscous and Ohmic damping, optically-thin radiation, thermal conduction, (2)
wave leakage across the sides of the loop boundaries, (3) wave leakage at the
chromospheric footpoints, (4) phase mixing in inhomogeneous loop regions, and
(5) resonant damping at the sides of loop boundaries. The first three effects 
are believed to be weak for fast kink-mode oscillations, while the latter two are
considered as most important (Goossens 1991; Poedts 2002; Ruderman \& Roberts 2002; 
Ofman \& Aschwanden 2002; Goossens, Andries, \& Aschwanden 2002; Erd\'elyi 2003).
First observational tests with TRACE data revealed that the scaling law of the
damping time as function of other physical parameters (loop length $L$ and period $P$)
favors the phase mixing mechanism (Ofman \& Aschwanden 2002), but the mechanism 
of resonant absorption can explain the observed damping times equally well if
the inhomogeneity length scale is a fraction of $\approx 15-50\%$ of the loop
radius (Goossens et al. 2002). More accurate tests to decide between these two
damping mechanisms require the knowledge of the inhomogeneity length scale $l$
and the density ratio $n_e/n_i$ between the external and internal electron density
of oscillating loops. The knowledge of the density ratio $n_e/n_i$ is also
required to calculate a coronal magnetic field strength $B$ from a loop with
oscillation period $P$ and length $L$, which is a fundamental tool of
{\sl coronal seismology} (Nakariakov \& Ofman 2001). 
In this paper here we measure for the first time these
additionally required parameters in 11 kink-mode oscillations events, for which 
the damping times have been reliably determined earlier (Aschwanden et al. 2002). 
This allows for a more rigorous quantitative test of the damping mechanisms, with
no free parameters for the theoretical model of resonant absorption. We find that
the mechanism of resonant absorption is commensurable with the observed damping times.
The data analysis and discussion of observational parameters
are discussed in Section 2, while conclusions are summarized in Section 3.

\section{   		DATA ANALYSIS 				}

We analyze 11 loop oscillation events from the study of Aschwanden et al. (2002) for
which a reliable damping time $t_D$ has been determined (e.g., see event \#1 in Fig.~1). 
The same data set of these
11 events is also studied in Ofman \& Aschwanden (2002) and in Goossens, Andries,
\& Aschwanden (2002). The date and times of the observations, the heliographic
coordinates, the inclination angles of the loop planes, the loop curvature radii, 
the oscillation periods, and the damping times are summarized in Table 1, 
extracted from Tables I and II in Aschwanden et al. (2002),
as well as one damping time from Nakariakov et al. (1999). 

\subsection{		Parameterization of Loop Skin Depth 		}

Damping of  oscillations and waves by
resonant absorption has been studied as a mechanism for coronal
heating. Most studies in this context are concerned with driven
waves. The interest for the present paper is in the eigenmodes
damped by resonant absorption. Hollweg and Yang (1988) derived an
analytical expression for the decay time in planar geometry for an
equilibrium model with a thin non-uniform boundary layer. They
translated their Cartesian result to cylindrical flux tubes and
were the first to point out that kink mode oscillations undergo
fast damping. In our view {\sl Hollweg decay} is a good name to refer to
this fast damping due to resonant absorption. Goossens et al. (1992)
derived  analytical expressions for the damping rate for 1-D
cylindrical flux tubes with thin non-uniform boundaries (TB) under
rather general conditions of the equilibrium magnetic field and
stationary equilibrium velocity field. Of particular relevance for
the present study is their Eq.~(77) for a static loop with a
straight field. It was derived under the assumption that the loop
is long so that the tube is thin (TT). Ruderman and Roberts (2002)
reconsidered the problem as an initial value problem. They arrived
at an analytical expression for the damping rate which is a
special case of the corresponding equation of Goossens et al.
(1992) (see their Eq.~56). Ruderman and Roberts then
specialized to a sinusoidal variation of density in the thin
nonuniform layer (their Eq.~71) and obtained the
corresponding decay time (their Eq.~73).
Their density profile $n(r)$ across a loop cross-section is parameterized by
\begin{equation}
	n(r) = \left\{ \begin{array}{ll}
		n_i  & \mbox{for $r < (a-l)$} \\
		n_i  \left[ {(1 + \chi )\over 2} - {(1 - \chi ) \over 2}
		\sin{ {\pi \over 2} {(2 r + l - 2 a) \over l }} \right] & \mbox{for $(a-l) < r < a$} \\
		n_e  & \mbox{for $r > a$}
	 	\end{array}
		\right.
\end{equation}
where ${\chi}=n_e/n_i$ is the density ratio of the external plasma $n_e$ to the
internal plasma density $n_i$ of the loop. The depth $l$ of the loop surface region 
might be called the {\sl ``skin depth''}, because it characterizes the depth of the 
outer envelope layer over which the density varies. So, $a$ is the outer loop radius,
$b=a-l$ the inner loop radius, and $R$ the mean radius, which defines also the
half width ($w_{loop}/2$) for the loop,
\begin{equation}
	R = {w_{loop}\over 2} = \left({a + b \over 2}\right) = a - {l\over 2} \ .
\end{equation}
So the density outside of the loop is $n(r>a)=n_e$, the skin region is bound 
by $b < r < a$, and the density inside this skin depth is $n(r<b)=n_i$. An example of
such a density profile is shown in Fig.~2 (bottom panel), for an inner radius
$b=0.4$ Mm and an outer radius $a=3.5$ Mm. 

With this parameterization, Ruderman \& Roberts (2002) derive a ratio of the 
exponential damping time $t_D$ to the oscillation period $P$ (their Eq.~73),
where we can replace $R \approx a$ in the thin-boundary approximation,
\begin{equation}
	\left({t_D \over P}\right)_{thin} 
	= {2 \over \pi} \left({a \over l}\right) {(1 + \chi ) \over (1 - \chi )} 
	\approx {2 \over \pi} \left({R \over l}\right) {(1 + \chi ) \over (1 - \chi )} \ ,
\end{equation}
\noindent Ruderman and Roberts (2002) obtained the above
analytical expression (and also their more general expression
(their Equation 56)) under the assumption that the non-uniform
layer is thin, meaning $l/a \ll 1.$ In the present paper we use the
Ruderman and Roberts formula to estimate the ratio $\chi$ for
loops with thick nonuniform layers. 
A generalization of this result for thick boundary layers thus
involves two corrections, one for the replacement of the outer
loop radius $a$ with the mean loop radius $R=a-l/2$, and a second one
that quantifies a correction factor $q_{TB}$ between the thick-boundary 
and thin-boundary treatment using the mean loop radius $R$. Thus,
the damping formula may be written in terms of $R$ and the
correction factor $q_{TB}$, 
\begin{equation}
	\left({t_D \over P}\right)_{thick} = q_{TB} \left({t_D \over P}\right)_{thin} 
	= q_{TB} \left( {2 \over \pi} {R \over l}\right) {(1 + \chi ) \over (1 - \chi )} \ .
\end{equation}
where the correction factor $q_{TB}$ depends on the boundary thickness ratio $(l/R)$ as
well as on the density ratio $\chi$, and has been calculated numerically
in Van Doorsselaere et al. (2003). For instance, for a density ratio of $\chi=1/3$,
the correction factor $q_{TB}$ varies in the range of $[0.75, 1.18]$.
In the fully non-uniform limit $l/R=2$, the correction value is 
$q_{TB}(l/R=2, \chi=1/3) \approx 0.75$. We use the numerically calculated values $q_{TB}$ in  
Table 2 to predict the external plasma density.

In a previous study we measured the oscillation periods $P$
and damping times $t_D$ of 11 events (Aschwanden et al. 2002). 
Here we attempt to measure the loop geometry
parameters $a$ and $b=a-l$, and the density ratio ${\chi}$ to test this
theoretical model (Eq.~4) of damping by resonant absorption. The {\sl density
contrast} of the oscillating loop is just the inverse ratio ${\chi}^{-1}=n_i/n_e$,
which is larger than unity for every detectable loop.

\subsection{		Loop Density Profiles					}

In order to measure cross-sectional density profiles $n(r)$ of coronal loops observed
in optically-thin EUV or soft X-ray emission, at least four effects play a role that
need to be taken into account: The subtraction of the background flux from the plasma 
in front and behind the oscillating loop along the line-of-sight (Fig.2),
the line-of-sight integration of the emission measure (Fig.3), 
the spatial smearing due to the transverse motion of an oscillating loop 
during an exposure time (Fig.4), and the point-spread function of the instrument (Fig.5). 

We start with the background subtraction, which is simply done by inspecting 
cross-sectional density profiles, averaged over some loop segment along the loop,
and by selecting the lowest flux values on both sides of the oscillating loop,
and interpolating a linear function between both sides (Fig.~2, top panel). 
Thus, the total EUV flux per pixel across a loop cross-section is defined by,
\begin{equation}
	F(r,t) = F_{back}(r,t) + F_{loop}(r,t) \ ,  
\end{equation}
which yields a time-averaged background flux $F_{back}$ (per pixel),
\begin{equation}
	F_{back} = < F_{back}(r,t) > \ , \qquad {r_{left}(t) < r < r_{right}(t)}
\end{equation}	
and a time-averaged loop flux $F_{loop}$ at the central axis of the loop,
\begin{equation}
	F_{loop} = max[ F_{loop}(r,t) ]  \ , \qquad {r_{left}(t) < r < r_{right}(t)}
\end{equation}	
where the loop boundaries $[r_{left}(t), r_{right}(t)]$ vary as function of time $t$
depending on the motion of the oscillating loop (see Fig.~6, left panel). 
The loop flux is related to the
electron density $n(z)$ along the line-of-sight $z$ by the emission measure $EM$ at 
pixel position $(x,y)$,
\begin{equation}
	{dEM(x,y,T) \over dT} = \int n^2(x,y,z,T) \ dz \ .
\end{equation}
The observed flux $F_{loop}(x,y)$ in a given filter specified by a temperature-dependent
instrumental response function $R(T)$ is obtained by integrating the emission measure 
$EM(T)$ with the response function $R(T)$ over all temperatures $T$,
\begin{equation}
	F_{loop}(x,y) = \int {dEM(x,y,T) \over dT} R(T) \ dT  \ .
\end{equation}
For an isothermal loop segment that is near-perpendicular to the line-of-sight,
we can obtain the radial flux profile $F(r)$ by integrating the density profile $n(r)$ 
specified in Eq.~1 along the line-of-sight $z$ (Fig.2 bottom and Fig.3),
\begin{equation}
	F_{loop}(r) = R(T) \int n^2(r'=\sqrt{r^2+z^2}) \ dz \ .
\end{equation}
The circular cross-section essentially causes a convolution of the density
profile $n(r)$ with a column depth $\Delta z$ that has a circular dependence 
$\Delta z \propto \sqrt{(a^2-r^2)}$, so that the flux profile $F(r)$ looks more 
gaussian-like (Fig.~2, middle panel) than the trapezoidal shape of the density function
(Fig.~2, third panel). We fit the (normalized) theoretical loop profile (Eq.~1) to the 
observed flux profiles $F_{loop}(r)$ by optimization of the parameters $a$ and $l$, 
using a {\sl Powell minimization method} (Press et al. 1986). 

The measured total flux at the center ($r=0$) of the loop is 
\begin{equation}
	F_{total}= \left[ n_i^2 w_{loop} + n_e^2 (z_{back}-w_{loop}) \right] R(T) \ ,
\end{equation}
with $n_i$ being the internal loop density, $w_{loop}$ the mean loop width (Eq.~2),
$n_e$ being the external or average background  
density extended over a column depth $z_{back}$. The mean background flux measured at the 
sides of the loop is,
\begin{equation}
	F_{back}= \left[ n_e^2 \ z_{back} \right] R(T) \ ,
\end{equation}
yielding a background-subtracted flux of
\begin{equation}
	F_{subtr} = F_{total} - F_{back} = \left[ (n_i^2 - n_e^2) w_{loop} \right] R(T) \ ,
\end{equation}
yields then the difference of the squared densities. Thus we cannot determine the internal
loop density $n_i$ directly, but only as function of the external density $n_e$,
\begin{equation}
	n_i = \sqrt{ { \Delta F_{subtr} \over w_{loop} R(T)} + n_e^2 } \ .
\end{equation}
In the limit of vanishing (background-subtracted) loop flux, $F_{subtr} \mapsto 0$,
the densities inside ($n_i$) and outside of the loop ($n_e$) become identical.
The density profile fitting is performed for every time step (typically
10-30 images) of the 11 oscillation events (e.g., see Fig.~6). The results of the best-fit parameters 
$a$, $l$ and $\sqrt{n_i^2-n_e^2}$, with the mean and standard deviation from averaging over 
all ($\approx$20-30) time steps, are given in Table 2.

The data analysis procedure is illustrated in Figs.1-8. Fig.2 shows a fit of a cross-section
profile $n(r,t_i)$ to the observed flux $F_{loop}(r,t=t_i)$ at a single time step $t_i$.
Fig.6 shows the fits as function of time $t_i, i=1,...,n$, and Fig.7 shows the variation
of the measured parameters as function of time, $F_{loop}(t)$ and $F_{back}(t)$ (Fig.7 top),
the oscillation amplitude $A(t)$ (Fig.7 middle), and the cross-section parameters $a(t)$, $l(t)$, and
$a(t)-l(t)$ (Fig.7 bottom), with the means and standard deviations indicated. Fig.8 shows a summary plot
of the average cross-section fits, for each of the 11 events. 

\subsection{		Oscillatory Motion Smearing		}

We have to be aware that every TRACE image has been recorded with a finite exposure time
of typically $\Delta t_{exp} \approx 5-10$ s. Since we are measuring the loop profiles
in perpendicular direction to the loop axis, the oscillatory motion of the transverse
kink mode introduces a smearing that transforms a rectangular density profile into
a trapezoidal profile (Fig.~4), if not corrected. For typical oscillation speeds of 
$v_{max} \approx 10-100$ km s$^{-1}$ (see Table 2 in Aschwanden et al. 2002) we expect
a motion of $\Delta r = v_{max} \ \Delta t_{exp} \approx 50-1000$ km, which corresponds
to $\Delta r \approx 0.1-2.8$ TRACE pixels with a pixel size of 0.5". 
However, the observed loop radii 
were found to be in the range of $a \approx 2000-12,000$ km, so at least an order
of magnitude larger. For most times the actual speed is smaller at an arbitrary phase
of the sinusoidal oscillations, $v(t) < v_{max}$.
We measured the actual amount of smearing for every fit and
found that it amounted indeed in all cases to a fraction of less than 0.05-0.1 of the loop width.
Therefore we neglected this effect in the fitting procedure. 

\subsection{		Point Spread Function			}

The instrumental point-spread function contributes to some broadening and
smoothing of observed density profiles, and thus could affect our inversion of
loop density profiles from observed flux profiles. The point-spread function of
TRACE has been investigated in 171 \ang\ EUV image fits using a {\sl blind iterative
deconvolution (BID)} procedure (Richard Nightingale, private communication).
The shape of the TRACE point-spread function was found to have the shape of a 
4-sided pyramid with a square-shaped base rotated by $45^0$ 
with respect to the CCD raster (Fig.~5).
The point-spread function falls off from a central pixel with value 1.0 to
0.33-0.36 in the next-neighbor pixels, and almost to 0.0 in the second-next
neighbor pixels, for a pixel size of 0.5". Thus, the average full-width of the 
pyramid shape is $FWHM = 2 \times 0.5" (1-0.345)/0.5 = 1.3"$.
 
Independently, the TRACE point-spread function has also been characterized with
a {\sl BID} procedure by Golub et al. (1999), who also found a slightly elongated
shape at a position angle of $45^0$, with a FWHM of 3 pixels in one direction and
2 pixels in the orthogonal direction, yielding a $FWHM=2.5 \times 0.5" = 1.25"$,
which is consistent with the former measurement. 

Since our measured loop widths ($w=2a-l \approx 2 ... 14$ Mm, see Table 1) are in the
average at least an order of magnitude larger than the FWHM of the point-spread 
function ($FWHM \approx 1.3 \times 0.725 = 0.94$ Mm, we neglect it in fitting of
the density profiles to the flux profiles. 

\subsection{		Predicted External Plasma Density	}

After we have measured the loop profile parameters $a$ and $l$ (Section 2.1 and Table 2),
and using the measurements of the observed loop oscillation periods $P$ and damping times
$t_D$ from the previous study (Aschwanden et al. 2002), we have only one free parameter
left in the damping time expression (Eq.~4),
namely the external-to-internal density ratio ${\chi}=n_e/n_i$ of the oscillating loop.
Because it is difficult to measure the ambient plasma density $n_e$
of an oscillating loop, we do not explicitly predict the damping time ratio $t_D/P$ based
on uncertain densities $n_e$ with relation (12), but rather do it the other way around by
using the theoretical relation (12) to predict the ambient density $n_e$, which can then
be compared with observational measurements. 

As Eq.~(4) shows, the shortest damping time ratios occur for a loop in vacuum, i.e. for
${\chi}=0$. We list these minimum ratios 
\begin{equation}
	\left( {t_D \over P} \right)_{min} = q_{TB} \left( {2 \over \pi} {R \over l}\right) 
\end{equation}
in Table 2, which are calculated from the measured values of $R=a-l/2$ and $l$
and using the fully non-uniform approximation $q_{TB}(l/R \approx 2) \approx 1.0$ (Van Doorsselaere et al. 2003). 
The resulting values of $(t_D/P)_{min}$ (Table 2; 7th column) reveal that they are all
lower than the observed values (Table 2; right-most column), as expected for ${\chi}>0$. This is
a first successful test of the theoretical model, in the sense that all 11 observed cases 
are able to provide a physical solution, namely a positive value for the density contrast, ${\chi}>0$.

In a next step we express the density contrast ${\chi}_D$ explicitly as function of the
other variables from Eq.~(4), (where the subscript in ${\chi}_D$ indicates here that it is derived
from the damping time $t_D$, instead of the standard definition in terms of density contrast,
${\chi}=n_e/n_i$, as defined in Eq.~1), 
\begin{equation}
	{\chi}_D = {(X - 1) \over (X + 1)} \ , \qquad 
	X={1 \over q_{TB}} {\pi \over 2} \left({l \over R}\right) \left({t_D \over P}\right) \ .
\end{equation}
We predict now the density contrasts ${\chi}_D$ based on the measured
ratios of damping times $t_D$ to periods $P$ ($(t_D/P)_{obs}$ 
in Table 2), using the fully non-uniform limit ($q_{TB}(l/R=2)=1.0$). 
We find values in the range of ${\chi}_D\approx 0.3-0.8$ (Table 3, 8th column).
Using the numerically calculated correction factors $q_{TB}(l/R)$ computed with the LEDA code
(Van Doorsselaere et al.~2003) for the actual observed values of $(l/R)$, indicated with 
${\chi}_{LEDA}$ in Table 3, we see that the approximation $q_{TB} \approx 1.0$ is a very
good approximation for fully-nonuniform loops (l/R=2).

\subsection{		Measurement of the External Plasma Density	}

In a next step in our analysis we attempt to estimate the external plasma density $n_e$ around the
oscillating loops from the measured background flux $F_{back}$ and loop position. 
The flux of the background is
composed of the emission measure along the line-of-sight in front and behind the oscillating loop.
We assume a stratified atmosphere  for the spatial and temporal average of the background flux, 
with an exponential density scale height ${\lambda}_T$ corresponding to a mean temperature $T$.
For the plasma detected in the TRACE 171 \ang\ passband, this mean temperature is $T \approx 1.0$ MK. 
Thus the vertical density profile of the detected coronal plasma is,
\begin{equation}
	n_e(h) = n_0 \exp{\left( - {h \over {\lambda}_T}\right)} \ ,
	\qquad {\lambda}_T = 47 \ {\rm Mm} \ \left({T \over 1.0\ {\rm MK}} \right) \ , 
\end{equation}
We need to calculate the column depth of a hydrostatically stratified
atmosphere along a line-of-sight as function of the distance $d$ from Sun center.
We define an {\sl equivalent column depth}, $z_{eq}(d,T)$, as function of the distance 
$d=r_{\odot}+h$ from Sun center, for a mean coronal temperature $T$,
\begin{equation}
        EM_{back}(d,T)=\int_{-\infty}^{\infty} n_e^2(h[z],T) dz = n_0^2(T) \ z_{eq}(d,T) \ .
\end{equation}
with $n_0(T)$ the coronal base density as defined in Eq.~(6).
From Eqs.~(6-7), the following relation follows for this equivalent column depth
(Table 3, 4th column),
\begin{equation}
        z_{eq}(d,T)=\int_{z_1}^{z_2} \exp \bigl( -{2
        \bigl[ \sqrt{d^2+z^2}-r_{\sun} \bigr] \over {\lambda}_T} \bigr) \ dz
        \quad {\rm for}\ d \ge r_{\odot} \ .
\end{equation}
with the integration limits $z_1=-\infty$ and $z_2=+\infty$ for above-the-limb locations
($d \ge r_{\odot}$).  Inside the disk ($d \le r_{\sun}$), we have only to change the
integration limit $z_2$ to,
\begin{equation}
        z_2(d)=-\sqrt{r_{\sun}^2-d^2} \quad {\rm for}\ 0 < d < r_{\odot} \ .
\end{equation}
The column depths $z_{eq}(d,T)$ are shown in Aschwanden \& Acton (2001; Fig.3 therein)
for a height range from disk center $(d=0)$ to one solar radius outside the limb 
($d=2 r_{\odot}$), for temperatures in the range of $T=1.0-4.0$ MK.
At disk center $(d=0)$, the equivalent column depth matches the
emission measure scale height, which is the half density scale height
$({\lambda}_{EM}={\lambda}_T/2)$.
At the limb ($d=r_{\odot}$), there is in principle a discontinuous change
by a factor of 2, which, however, is difficult to measure because
of the extremely high instrumental resolution required to resolve
this jump. Above the limb, the column depth drops quickly with height. 

We can now relate the observed background flux $F_{back}(d)$, measured at a line-of-sight with
distance $d$ from disk center, to the emission measure $EM_{back}(d)$,
using the instrumental response function $R(T)$,
\begin{equation}
	F_{back}(d) = EM_{back}(d) R(T) \ .
\end{equation}
and determine the coronal base density $n_0$ (with Eqs.~9 and 10),
\begin{equation}
	n_0 = \sqrt{ {F_{back}(d) \over R(T) z_{eq}(d,T) } } \ .
\end{equation}
To estimate the ambient density around the oscillating loop in height $h_{osc}$,
we have to use the hydrostatic model of Eq.~(6),
\begin{equation}
	n_e(h=h_{osc}) = n_0 \exp{\left( - {h_{osc} \over {\lambda}_T}\right)} \ ,
\end{equation}
Obviously we need measurements of the center-limb distance $d$ of the location 
of the oscillating loop segment, as well as an estimate of the altitude $h_{osc}$ 
of the oscillating loop segment. In the previous study (Aschwanden et al. 2002)
we measured the heliographic longitude difference $(l_1-l_0)$ and latitude difference
$(b_1-b_0)$ of the midpoint of the loop baseline to disk center, the inclination
angle $\vartheta$ of the loop plane, and the loop curvature radius $R_{curv}$.
We list these parameters of our 11 analyzed loops in Table 1. With these 
geometrical parameters we can now determine the projected distance $d$ of the location
of the oscillating loop segment to disk center (Table 3, 3rd column),
\begin{equation}
	d = ( r_{\odot} + h_{osc} )
		\sin{\left[ (l_1-l_0)^2 + (b_1-b_0)^2 \right]} \ , 
\end{equation}
and the height $h_{osc}$ of the oscillating loop segment above the solar surface
(Table 3, 2nd column),
\begin{equation}
	h_{osc} = R_{curv} \cos \vartheta  \ .
\end{equation}
Inserting these parameters $d$ and $h_{osc}$ into Eqs.(6)-(12) 
we obtain now an estimate of the ambient density at the height of the oscillating loops, 
$n_e(h=h_{osc})$. The so evaluated density values $n_e$ are given in Table 3 (5th column), 
along with the inferred internal densities $n_i$ (Eq.~14) 
and density ratios ${\chi}={n_e/n_i}$. 
These density ratios can now be compared with
the predicted density contrast ${\chi}_D$ (Eq.~5) from the observed damping times
(Table 3). The uncertainties of the derived 
parameters were estimated according to the error propagation law (Appendix A). 
We find that density ratio is consistent with the model of resonant absorption within 
a factor of ${\chi}_{LEDA}/\chi \approx 1.2-3.5$ (Table 3, right-most column,
excluding the lowest and highest extreme value).

\subsection{Temperature Corrections}

In our analysis we used the temperature of $T=1.0$ MK that corresponds to the peak of
the TRACE 171 \ang\ passband, in which all the analyzed oscillating loops were detected.
This peak temperature is certainly representative for the background plasma along the
line-of-sight, because it represents an average over many coronal structures which are
detected in a given passband with the highest probability near the peak temperature of
the temperature sensitivity. So the temperature should not affect any derived parameter
based on the background plasma, such as the external plasma density $n_e$. 

What's about the temperature inside the oscillating loops. 
Since the FWHM of the temperature response function in 171 \ang\ is about $\Delta T_{171}/T_{171}
=(1.2-0.8)/1.0=0.4$, a loop is detected with a probability of 67\% in this temperature range.  
The peak response function we used is $R_{171}(T=1.0 MK) = 1.1 \times 10^{-26}$ DN s$^{-1}$ cm$^{5}$
(see e.g. Fig.~12 in Aschwanden et al. 2000). If we approximate the response function with a 
gaussian curve, single-temperature plasma structures are detected with a probability of 24\% at
a sensitivity that is less than 50\% of the peak response. So, statistically, in every 4th loop
we may have overestimated the response function by a factor of $\gapprox 2$, which is equivalent
to an underestimation of the true loop density by a factor of $\lapprox \sqrt{2}$. Therefore
the resulting density contrast ${\chi}=n_e/n_i$ could be a factor of 1.4 higher for every 4th
loop. In the statistical average, however, this temperature correction is not sufficient to
explain the average discrepancy between the density ratios, i.e., ${\chi}_{LEDA}/{\chi} \approx 
1.2-3.5$.

\subsection{Coronal Filling Factors}

Another not considered effect is the spatial filling factor, which affects both the plasma
determination external and internal to the oscillating loops. Generally, if the filling factor
is less than unity, density derivations from the emission measure (Eqs.~8, 11, 19) result into
an underestimate of the density. If both the internal and external plasma is subject to the
same filling factor, this effect would cancel out in the density contrast ${\chi}=n_e/n_i$
and no correction is needed. However, we think that the oscillating fluxtubes, especially
those with small diameters are more likely to be solidly filled with plasma than the wide
bundles of fluxtubes, or the background corona. Inquiring the diameters of the oscillating
loops we find large radii $a \gapprox 10$ Mm only for two cases (3a and 16a in Table 2),
which show the same discrepancy between ${\chi}$ and ${\chi}_D$ as the other cases (Table 3), 
so a correction by a filling factor of loops cannot improve the consistency between data 
and model either.

On the other side we can ask whether the filling factor of the background corona has an
effect on our model. With our stratified coronal model we applied for the temporal and
spatial average (Eq.~6), we inferred a density contrast of ${\chi}=0.30\pm0.16$.
If the background
corona is subject to a filling factor less than unity, the true ambient density around a
loop could be lower or higher. A possible bias towards a higher value could result in 
active regions, where high-density concentrations are more likely around oscillating loops.
If we consider such a filling factor bias and assume that the ambient density
around an oscillating loop in an active region is actually higher, the density contrast value
${\chi}=n_e/n_i$ increases. The mismatch is in the average ${\chi}_{LEDA}/\chi \approx 1.2-3.5$, 
which could be reconciled with 
correspondingly higher ambient densities around the oscillating loops. This higher
density does not necessarily need to be plasma with a temperature of $T\approx 0.8-1.2$ MK
as detected with TRACE 171 \ang\ , it could be plasma of higher temperature, say in the
range of $T\approx 1.2-2.0$ MK, as many differential emission measure distributions
inferred in active regions suggest (e.g. Brosius et al. 1996; Aschwanden \& Acton 2001).
Improvements in the determination of the mean external density $n_e$ therefore require
the knowledge of the differential emission measure function, which demands multi-filter
data.

\subsection{Comparison with Phase Mixing Model}

The theory for damping due to resonant absorption for thin non-uniform layers
predicts that the damping time $t_D$ is a function of the period $P$, the
geometric ratio $(R/l)$, and the external/internal density ratio ${\chi}=n_e/n_i$ 
(Eq.~4), without any free parameter, 
\begin{equation}
	t_D^{RA} = {2\ q_{TB}\ R \ P \over \pi \ l} {(1 + \chi ) \over (1 - \chi)} \ .
\end{equation}
We test this scaling law in Fig.~9 (right frame) and find a mean ratio of
$t_D^{RA}/t_D^{obs} = 0.37 \pm 0.15$, so theory and observations agree within
a factor of 3, where the discrepancy probably is due to the underestimation 
of the background density when measured with a narrow-band filter. 

Alternatively, we might test the scaling law for phase mixing (Heyvaerts \& Priest 1983;
Roberts 2000), which predicts the following dependence,
\begin{equation}
	t_D^{PM} = \left( { 6 L^2 l_{inh}^2 \over \nu \pi^2 {\rm v}_A^2 }\right)^{1/3}
		= \left( { 3 \over \nu \pi^2} \right)^{1/3} ( P l )^{2/3}  
\end{equation}
where $L$ is the loop length, $l_{inh}$ the scale of inhomogeneity (which we set equal to
our skin depth here), $\nu$ is the coronal viscosity, v$_A$ the Alfv\'en speed inside
the fluxtube, which amounts to v$_A = \sqrt{2} L/P$ for the kink mode in a low-$\beta$
plasma (Nakariakov \& Ofman 2001). We calculate the predicted damping times with a standard
value of the coronal viscosity, $\nu = 4 \times 10^{13}$ cm$^2$ s$^{-1}$, and
plot them in Fig.9 (left frame). We find an average ratio of 
$t_D^{PM}/t_D^{obs} = 0.79 \pm 0.19$, which closely agrees with the observations.
Thus the model of phase mixing cannot be excluded as alternative interpretation. 

A corresponding test of the scaling law $t_D \propto P$ 
for resonant absorption and $t_D \propto (L P)^{2/3}$ has been performed in
Ofman \& Aschwanden (2002) that showed also a slight preference for phase mixing. 
The test here, however, is more constrained.
There are three differences to the former study: (1) we do not
make the assumption that the spatial scale of inhomogeneity $l_{inh}$ is proportional to
the loop length $L$ or loop width $w$; 
(2) The loop widths $w_{loop}=a+b$ are measured here from the deconvolved
density profiles and not from the FWHM of the flux profiles, and (3) we measure here
the (outer) loop radius $a$ and spatial scale of inhomogeneity $l$ separately, which were
set equal to each other in the former study. Nevertheless, we obtain similar results
that both models are roughly consistent with the observations.

\section{CONCLUSIONS}

In this study we modeled the cross-sectional density profiles $n_e(r)$ of oscillating loops,
specified by the outer radius $a$, skin depth $l$, internal density $n_i$, and
external density $n_e$. These parameters allow us to test the theoretically predicted
relation between the damping time $t_D$, oscillation period $P$, geometry
($a,l$), and density parameters ($n_e, n_i$) for the damping mechanism of resonant
absorption. Because we can measure all these observables we have no free parameters
in the model and thus are able to perform a very strict consistency test between
theory and observations. The alternative damping mechanism of phase mixing
can be tested with these measured parameters also, but there is a free parameter,
namely the viscosity, which cannot directly be constrained by observations to date.
Our observational test yields the following results:

\begin{enumerate}

\item{The means and standard deviations of our measured parameters are (see Table 4):
	Outer loop radius $a=4.5\pm3.5$ Mm, loop skin depth $l=3.9\pm3.1$ Mm,
	skin depth ratio $l/a=0.85\pm0.08$, 
	internal loop density $n_i=(1.4\pm0.7)\times 10^9$ cm$^{-3}$,
	external loop density $n_e=(0.36\pm0.18)\times 10^9$ cm$^{-3}$,
	density ratio $\chi=n_e/n_i=0.30\pm0.16$. These are the averages of
	11 oscillating loops events.}

\item{In a previous study we measured the corresponding oscillation
	periods, $P=317\pm114$ s, the damping times $t_D=574\pm320$ s,
	which yield a mean ratio of $t_D/P=1.8\pm0.8$. According to the resonant
	damping model of Rudermann \& Roberts (2002), originally derived for the
	thin-boundary approximation and now generalized for the thick-boundary
	approximation by Van Doorsselaere et al. (2003), the 
	observed damping times constrain (under the assumption of damping by
	resonant absorption) a density ratio of ${\chi}_D=0.53\pm0.12$,
	which is a factor of ${\chi}_{LEDA}/\chi \approx 1.2-3.5$ higher than that
	measured from the background
	fluxes with the TRACE 171 \ang\ filter. It is likely that this discrepancy 
	factor results from the neglected hotter plasma with $T\gapprox 1.5$ MK
	that is not detected with the 171 \ang\ filter. With this correction,
	the damping model or resonant absorption can be considered as a successful
	theory to explain the observed damping times. Alternatively, the model of
	phase mixing is also found to be consistent with the data.}

\item{The damping model by resonant absorption provides a direct diagnostic of the
	density ratio ${\chi}_D=n_e/n_i$. The observed parameters of the loop cross-section
	profiles vary very little, the ratio $a/l=1.18\pm0.11$ varies only by $\approx 10\%$
	and can be neglected in the damping formula. The correction factor for the
	thick-boundary treatment can then be taken in the fully non-uniform limit,
	$q_{TB}(l/R=2)\approx 1.0$. Therefore we have a very simple
	relation that predicts the number of oscillations $N_{osc}=t_d/P$ as function
	of the density ratio ${\chi}_D=n_e/n_i$, or vice versa (Eq.~16),
	\begin{equation}
		N_{osc} = {t_D \over P} \approx {1 \over \pi} {1 + {\chi}_D \over 1 - {\chi}_D} \ ,
	\end{equation} 	
	\begin{equation}
		{\chi}_D = {n_e \over n_i} \approx {\pi N_{osc} - 1 \over \pi N_{osc} + 1} \ .
	\end{equation} 	
	The relation is plotted in Fig.~10 and can be used as an efficient density 
	diagnostic.}

\end{enumerate}

This study provides new support for the interpretation of damping mechanism of coronal
loop oscillations in terms of the resonant absorption process. A new effect we learned
from this study is the sensitivity to the density contrast between the loop and the
ambient plasma. In vacuum, the loop oscillations would be damped within a half
oscillation period, $t_D/P \approx 0.5$. However, the higher the ambient plasma
density is, the less severe is damping by resonant absorption, so that undamped
oscillations can only be supported if the density contrast is very little. Asking
the question why only a small subset of all active region loops exhibit oscillations
after a global triggering event, e.g. during a flare or a filament destabilization
(Schrijver et al. 2002), the mechanism of resonant absorption provides a plausible
explanation that oscillations are most favored in loops with little density constrast to
the ambient plasma, while all other loops with a large density contrast are strongly
damped within a half oscillation period. For future work to study the role of resonant 
absorption we recommend to model the differential emission measure distribution of the
coronal plasma with multi-filter data to obtain a better estimate of the coronal
background density.

\subsection*{Acknowledgements:}
We thank Prof. Bernie Roberts, Valery Nakariakov, Bart DePontieu, and Karel Schrijver 
for helpful discussions. Part of this work was supported by NASA contract NAS5-38099 (TRACE).

\subsection*{Appendix A: Estimates of Parameter Uncertainties}

The variables of the damping time $t_D$ and period $P$ (Table 1) have been determined
in Aschwanden et al. (2002) without an estimate of the uncertainty. Based on multiple
trials with different background subtractions we estimate the uncertainty of the damping time 
to be of order ${\sigma}_{t_D} \approx t_D/3$, e.g., compare the result of $t_D=1200$ s
for event 1a) in Aschwanden et al. (2002) versus $t_D=870$ s in Nakariakov et al. (1999).
The error in the period measurement $P$ can be neglected because repeated fitting with different
background subtractions reproduced this value within a few percent, so the error is much 
smaller than the error of the damping time $t_D$. Also the errors in the parameters
$l_1-l_0, b_1-b_0, R_{curv}, \vartheta$ and the derived quantities $h_{osc}$ (Eq.~25) and  
$d$ (Eq.~24) are accurate to a few percent and thus the uncertainties can be neglected.
The equivalent column depth $z_{eq}$ (Eq.19) is a theoretical quantity that has no measurement
error.

For all parameters directly measured in this study, $F_{loop}$, $F_{back}$, $a$, and $l$ (Table 2), 
we determined the uncertainties ${\sigma}_{F_{loop}}$, ${\sigma}_{F_{back}}$, ${\sigma}_{a}$, 
and ${\sigma}_{l}$ from the standard deviations that resulted 
by averaging the fits of all times per event, $t_i, i=1,...,n$, with typically $n\approx 20-30$ 
time steps per event. 

The uncertainties of the derived parameters $n_i$ (Eq.14) $q_D=t_D/P$ (Eq.4), $X$ (Eq.6), ${\chi}_D$ (Eq.6),
$n_0$ (Eq.~22), $n_e$ (Eq.~23), and $\chi = n_e/n_i$ (after Eq.~1) were calculated with the error propagation law, 	
$$
	{\sigma}_{n_i} \approx n_i \left( {1 \over 2 F_{loop}} \right) {\sigma}_{F_{loop}} \ ,
	\eqno(A1)
$$
$$
	{\sigma}_{q_D} = 
	{q_D}\sqrt{ ({\sigma}_a/a)^2 + ({\sigma}_l/l)^2 } \ ,
	\eqno(A2)
$$
$$
	{\sigma}_X = X \sqrt{ ({\sigma}_{t_D}/t_D)^2 + ({\sigma}_a/a)^2 + ({\sigma}_l/l)^2 } \ ,
	\eqno(A3)
$$
$$
	{\sigma}_{{\chi}_D} = {\chi}_D \left( {2 \over X^2-1} \right) {\sigma}_X \ ,
	\eqno(A4)
$$
$$
	{\sigma}_{n_0} = n_0 \left( {1 \over 2 F_{back}} \right) {\sigma}_{F_{back}} \ ,
	\eqno(A5)
$$
$$
	{\sigma}_{n_e} = n_e \left( {1 \over 2 F_{back}} \right) {\sigma}_{F_{back}} \ ,
	\eqno(A6)
$$
$$
	{\sigma}_{\chi} = {\chi} \sqrt{ ({\sigma}_{n_e} / n_e)^2 + ({\sigma}_{n_i} / n_i)^2} \ .
	\eqno(A7)
$$

\clearpage

\subsection*{References} 

\refer{Aschwanden,M.J., Tarbell,T., Nightingale,R., Schrijver,C.J., Title,A., Kankelborg,C.C., 
	Martens,P.C.H., and Warren,H.P. 2000, ApJ 535, 1047.}
\refer{Aschwanden,M.J., \& Acton,L.W. 2001, ApJ 550, 475.}
\refer{Aschwanden,M.J., DePontieu,B., Schrijver,C.J., and Title,A.M. 2002, SP 206, 99.}
\refer{Aschwanden,M.J., Schrijver,C.J., Winebarger,A.R., and Warren,H.P. 2003, ApJ 588, L49.}
\refer{Aschwanden,M.J.  2003, in "Turbulence, Waves, and Instabilities in the Solar Plasma", 
	NATO Advanced Research Workshops, 16-20 Sept 2002, Budapest, Hungary, 
	(eds. R. von Fay-Siebenburgen, K. Petrovay, B. Roberts, and M.J. Aschwanden).}
\refer{Brosius,J.W., Davila,J.M., Thomas,R.J., and Monsignori-Fossi,B.C.
 	1996, ApJS 106, 143.}
\refer{Erd\'elyi,R. 2003, in "Turbulence, Waves, and Instabilities in the Solar Plasma", 
	NATO Advanced Research Workshops, 16-20 Sept 2002, Budapest, Hungary, 
	(eds. R. von Fay-Siebenburgen, K. Petrovay, B. Roberts, and M.J. Aschwanden).}
\refer{Feldman,U. 1992, Physica Scripta 46, 202.}
\refer{Golub,L. and 11 co-authors 1999, Physics of Plasmas 6, No.5, p.2205.}
\refer{Goossens,M. 1991, {\sl Magnetohydrodynamic waves and wave heating in nonuniform
	plasmas}, in E.R.Priest and A.W.Hood (eds.): {\sl Advances in Solar System
	Magnetohydrodynamics}, Cambridge: Cambridge University Press, p.137.}
\refer{Goossens,M., Andries,J., \& Aschwanden,M.J. 2002, AA 394, L39.}
\refer{Heyvaerts,J. \& Priest,E.R. 1983, AA 117, 220.}
\refer{Hollweg,J.V. \& Yang,G. 1988, J.~Geophys.Res., 93, 5423.}
\refer{Nakariakov,V.M., Ofman,L., DeLuca,E., Roberts,B., Davila,J.M.
 	1999, Science 285, 862.}
\refer{Nakariakov,V.M. \& Ofman,L. 2001, AA 372, L53.}
\refer{Ofman,L. \& Aschwanden,M.J. 2002, ApJ 576, L153.}
\refer{Poedts,S. 2002, in {\sl Proc. SOLMAG: Magnetic Coupling of the Solar Atmosphere
	Euroconference and IAU Colloq. 188}, ESA SP-505 (ed. H. Sawaya-Lacoste),
	Santorini, p. 273.}
\refer{Press,W.H., Flannery,B.P., Teukolsky,S.A., and Vetterling,W.T.
 	1986, {\sl Numerical Recipes, The Art of Scientific Computing}, 
	Cambridge: Cambridge University Press.}
\refer{Roberts,B., Edwin,P.M., \& Benz,A.O. 1984, ApJ 279, 857.}
\refer{Roberts,B., 2000, SP 193, 139.}
\refer{Roberts,B. \& Nakariakov,V. 2003, in "Turbulence, Waves, and Instabilities in the Solar Plasma", 
	NATO Advanced Research Workshops, 16-20 Sept 2002, Budapest, Hungary, 
	(eds. R. von Fay-Siebenburgen, K. Petrovay, B. Roberts, and M.J. Aschwanden).}
\refer{Ruderman,M.S. \& Roberts,B. 2002, ApJ 577, 475.}
\refer{Schrijver,C.J., Aschwanden,M.J., and Title,A.M. 2002, SP 206, 69.}
\refer{Van Doorsselaere,T., Andries,J., Poedts,S., and Goossens,M. 2003, in preparation.}
\refer{White,S.M., Thomas,R.J., Brosius,J.W., \& kundu,M.R. 2000, ApJ 534, L203.}

\clearpage

\begin{deluxetable}{rrrrrrrrr}
\footnotesize
\tablecaption{Times, locations, loop geometries, oscillation periods, and damping times
	of 11 oscillation events analyzed in Aschwanden et al. (2002).}
\tablewidth{0pt}
\tablehead{
\colhead{No.}&
\colhead{Date and Time}&
\colhead{Heliographic}&
\colhead{Heliographic}&
\colhead{Loop}&
\colhead{Loop}&
\colhead{Oscillation}&
\colhead{Damping}\\
\colhead{}&
\colhead{of Observation}&
\colhead{Longitude}&
\colhead{Latitude}&
\colhead{Inclination}&
\colhead{Curvature}&
\colhead{Period}&
\colhead{Time}\\
\colhead{}&
\colhead{}&
\colhead{$l_0-l_{\odot}$}&
\colhead{$b_0-b_{\odot}$}&
\colhead{$\vartheta$}&
\colhead{$R_{curv}$}&
\colhead{$P$}&
\colhead{$t_D$}\\
\colhead{}&
\colhead{}&
\colhead{(deg)}&
\colhead{(deg)}&
\colhead{(deg)}&
\colhead{(Mm)}&
\colhead{(s)}&
\colhead{(s)}}
\startdata
 1a) & 1998-Jul-14 1259:57 & -15.6 & -27.6 &   7.0 &  47.0 &   261 &   870\tablenotemark{a} \\
 1b) & 1998-Jul-14 1257:38 & -15.5 & -26.0 &  19.0 &  24.0 &   265 &   300 \\
 1d) & 1998-Jul-14 1257:36 & -19.5 & -24.5 & -35.0 &  55.0 &   316 &   500 \\
 1f) & 1998-Jul-14 1256:32 & -19.6 & -24.5 & -44.0 &  57.0 &   277 &   400 \\
 1g) & 1998-Jul-14 1302:26 & -19.2 & -22.7 &  47.0 &  45.0 &   272 &   849 \\
 3a) & 1998-Nov-23 0635:57 &  82.3 & -27.7 & -12.0 &  99.0 &   522 &  1200 \\
 4a) & 1999-Jul-04 0833:17 &  26.0 & -27.3 & -14.0 &  74.0 &   435 &   600 \\
 5c) & 1999-Oct-25 0628:56 & -22.9 & -21.3 &   2.0 &  53.0 &   143 &   200 \\
10a) & 2001-Mar-21 0232:44 &  72.6 &  -3.8 &  20.0 &  77.0 &   423 &   800 \\
16a) & 2001-May-15 0257:00 &  22.7 & -18.3 &  39.0 &  68.0 &   185 &   200 \\
17a) & 2001-Jun-15 0632:29 & -48.7 & -28.0 &  41.0 &  33.0 &   396 &   400 \\
\enddata
\tablenotetext{a}{This value of $t_D=870$ s was measured in Nakariakov et al. (1999) and is
	also used in the study of Ofman \& Aschwanden (2002). An alternative value of
	$t_D=1200$ s was determined in Aschwanden et al. (2002). The difference reflects
	a typical uncertainty in the determination of the damping time $t_D$.}
\end{deluxetable}

\begin{deluxetable}{rrrrrrrr}
\footnotesize{\scriptsize\small\tiny}
\tablecaption{Best-fit parameters of loop cross-section fits to the same 11 events specified in Table 1.}
\tablehead{
\colhead{No.}&
\colhead{Loop flux}&
\colhead{Background}&
\colhead{Loop radius}&
\colhead{Skin depth}&
\colhead{Loop density}&
\colhead{Minimum ratio}&
\colhead{Observed ratio}\\
\colhead{}&
\colhead{$F_{subtr}$ [DN/s]}&
\colhead{$F_{back}$ [DN/s]}&
\colhead{$a$ [Mm]}&
\colhead{$l$ [Mm]}&
\colhead{$\sqrt{n_i^2-n_e^2}$ [$10^8$ cm$^{-3}$]}&
\colhead{$(t_D/P)_{min}$}&
\colhead{$(t_D/P)_{obs}$}}
\startdata
 1a) &   9.9$\pm$  1.6 &  23.2$\pm$  1.8 &   3.5$\pm$  0.4 &   3.3$\pm$  0.5 &  15.6$\pm$  1.3 &  0.35$\pm$ 0.53 &  3.33$\pm$ 1.11 \\
 1b) &  20.9$\pm$  5.0 &  28.0$\pm$  8.1 &   3.4$\pm$  0.5 &   2.9$\pm$  0.5 &  22.4$\pm$  2.7 &  0.41$\pm$ 0.26 &  1.13$\pm$ 0.38 \\
 1d) &   4.0$\pm$  3.1 &  18.6$\pm$  7.5 &   2.6$\pm$  0.9 &   2.1$\pm$  1.1 &  10.8$\pm$  4.3 &  0.48$\pm$ 0.59 &  1.58$\pm$ 0.53 \\
 1f) &   0.5$\pm$  0.3 &   7.3$\pm$  3.0 &   2.0$\pm$  0.5 &   1.6$\pm$  0.6 &   4.5$\pm$  1.6 &  0.49$\pm$ 0.42 &  1.44$\pm$ 0.48 \\
 1g) &   7.0$\pm$  2.8 &  43.2$\pm$  3.7 &   3.4$\pm$  0.7 &   2.9$\pm$  1.0 &  12.7$\pm$  2.8 &  0.44$\pm$ 0.55 &  3.12$\pm$ 1.04 \\
 3a) &  12.7$\pm$  5.3 &  34.7$\pm$  1.3 &  12.4$\pm$  4.1 &  10.8$\pm$  4.3 &   9.1$\pm$  2.0 &  0.41$\pm$ 0.40 &  2.30$\pm$ 0.77 \\
 4a) &  12.6$\pm$  2.2 &  54.4$\pm$ 12.6 &   2.8$\pm$  0.5 &   2.7$\pm$  0.5 &  19.8$\pm$  1.7 &  0.34$\pm$ 0.38 &  1.38$\pm$ 0.46 \\
 5c) &  13.9$\pm$  2.9 &  44.8$\pm$  6.3 &   2.5$\pm$  0.2 &   2.3$\pm$  0.2 &  21.9$\pm$  2.3 &  0.37$\pm$ 0.26 &  1.40$\pm$ 0.47 \\
10a) &  25.9$\pm$  8.4 &  40.5$\pm$  6.5 &   4.6$\pm$  0.5 &   3.6$\pm$  0.8 &  20.7$\pm$  3.4 &  0.48$\pm$ 0.36 &  1.89$\pm$ 0.63 \\
16a) &   2.5$\pm$  0.5 &   6.3$\pm$  1.5 &  10.2$\pm$  2.9 &   9.0$\pm$  3.3 &   4.4$\pm$  0.4 &  0.40$\pm$ 0.59 &  1.08$\pm$ 0.36 \\
17a) &   1.2$\pm$  0.7 &  53.6$\pm$  2.0 &   1.6$\pm$  0.6 &   1.2$\pm$  0.7 &   7.1$\pm$  2.9 &  0.57$\pm$ 0.40 &  1.01$\pm$ 0.34 \\
\enddata
\end{deluxetable}

\clearpage

\begin{deluxetable}{cccccccccc}
\setlength{\tabcolsep}{0.04in}
\footnotesize{\scriptsize\small\tiny}
\tablecaption{Model parameters of external background plasma density and density contrast.}
\tablehead{
\colhead{No.}&
\colhead{Loop apex}&
\colhead{Distance}&
\colhead{Equivalent}&
\colhead{External}&
\colhead{Internal}&
\colhead{Density}&
\colhead{Predicted}&
\colhead{Numerical}&
\colhead{Density}\\
\colhead{No.}&
\colhead{altitude}&
\colhead{from Center}&
\colhead{Column depth}&
\colhead{density}&
\colhead{density}&
\colhead{ratio}&
\colhead{ratio}&
\colhead{ratio}&
\colhead{ratio}\\
\colhead{}&
\colhead{$h_{osc}$ [Mm]}&
\colhead{$d_{osc}/r_{\odot}$}&
\colhead{$z_{eq}/r_{\odot}$}&
\colhead{$n_e$ [$10^8$ cm$^{-3}$]}&
\colhead{$n_i$ [$10^8$ cm$^{-3}$]}&
\colhead{${\chi}$}&
\colhead{${\chi}_D$}&
\colhead{${\chi}_{LEDA}$}&
\colhead{${\chi}_{LEDA}/{\chi}$}}
\startdata
 1a) &  46.6 &  0.561 &  0.040 &   3.2$\pm$  0.3 &  15.9$\pm$  1.3 &  0.20$\pm$ 0.02 &  0.81$\pm$ 0.27 &  0.70 &  3.5 \\ 
 1b) &  22.7 &  0.521 &  0.039 &   6.0$\pm$  0.7 &  23.2$\pm$  2.7 &  0.26$\pm$ 0.04 &  0.46$\pm$ 0.28 &  0.43 &  1.6 \\ 
 1d) &  45.1 &  0.553 &  0.040 &   3.0$\pm$  1.1 &  11.2$\pm$  4.3 &  0.27$\pm$ 0.14 &  0.53$\pm$ 0.46 &  0.52 &  2.0 \\ 
 1f) &  41.0 &  0.551 &  0.040 &   2.1$\pm$  0.7 &   4.9$\pm$  1.6 &  0.42$\pm$ 0.19 &  0.50$\pm$ 0.35 &  0.49 &  1.2 \\ 
 1g) &  30.7 &  0.518 &  0.039 &   6.3$\pm$  1.3 &  14.2$\pm$  2.8 &  0.44$\pm$ 0.13 &  0.75$\pm$ 0.28 &  0.70 &  1.6 \\ 
 3a) &  96.8 &  1.137 &  0.008 &   3.1$\pm$  0.6 &   9.6$\pm$  2.0 &  0.32$\pm$ 0.09 &  0.70$\pm$ 0.26 &  0.64 &  2.0 \\ 
 4a) &  71.8 &  0.675 &  0.044 &   2.7$\pm$  0.2 &  20.0$\pm$  1.7 &  0.14$\pm$ 0.02 &  0.60$\pm$ 0.37 &  0.47 &  3.4 \\ 
 5c) &  53.0 &  0.559 &  0.040 &   3.9$\pm$  0.4 &  22.3$\pm$  2.3 &  0.18$\pm$ 0.03 &  0.58$\pm$ 0.25 &  0.50 &  2.8 \\ 
10a) &  72.4 &  1.054 &  0.109 &   1.5$\pm$  0.2 &  20.8$\pm$  3.4 &  0.07$\pm$ 0.02 &  0.60$\pm$ 0.27 &  0.58 &  8.1 \\ 
16a) &  52.8 &  0.524 &  0.039 &   1.5$\pm$  0.1 &   4.7$\pm$  0.4 &  0.32$\pm$ 0.04 &  0.46$\pm$ 0.60 &  0.41 &  1.3 \\ 
17a) &  24.9 &  0.860 &  0.061 &   6.3$\pm$  1.9 &   9.5$\pm$  2.9 &  0.66$\pm$ 0.28 &  0.28$\pm$ 0.36 &  0.36 &  0.5 \\ 
\enddata
\end{deluxetable}

\clearpage

\begin{deluxetable}{lr}
\footnotesize{\scriptsize\tiny}
\tablecaption{Means and standard deviations of measured parameters in 11 oscillation events.}
\tablehead{
\colhead{Parameter}&
\colhead{Mean and standard deviation}}
\startdata
Loop curvature radii $R_{curv} =$		& $57\pm21$ Mm \\
Oscillation period $P =$			& $317 \pm 114$ s \\
Damping time $t_D =$				& $574 \pm 320$ s \\
Observed number of oscillations $t_D/P =$	& $1.8\pm0.8$ \\
Predicted minimum of ratio $(t_D/P)_{min}=$	& $0.32\pm0.05$ \\
Outer loop radius $a=$				& $4.5\pm3.5$ Mm \\
Inner loop radius $a-l=$			& $0.6\pm0.5$ Mm \\
Mean loop width $w_{loop}=2a-l=$		& $5.1\pm3.9$ Mm \\
Loop skin depth $l=$				& $3.9\pm3.1$ Mm \\
Relative loop skin depth $l/R=$			& $1.5\pm0.2$ \\
Loop density $n_i=$				& $1.4\pm0.7\ 10^9$ cm$^{-3}$ \\
External plasma density $n_e(T=1 MK)=$		& $0.36\pm0.18\ 10^9$ cm$^{-3}$ \\
Predicted external plasma density $n_e=n_i {\chi}_{LEDA}=$ & $0.76\pm0.36\ 10^9$ cm$^{-3}$ \\
Density ratio $\chi=n_e(T=1 MK)/n_i=$		& $0.30\pm0.16$ \\
Predicted density ratio ${\chi}_{LEDA}=n_e/n_i=$& $0.53\pm0.12$ \\
Prediction ratio $n_e/n_e(T=1 MK)=\chi_{LEDA}/{\chi}=$ & $2.5\pm2.1$ \\
\enddata
\end{deluxetable}


\clearpage

\begin{figure} 
\plotone{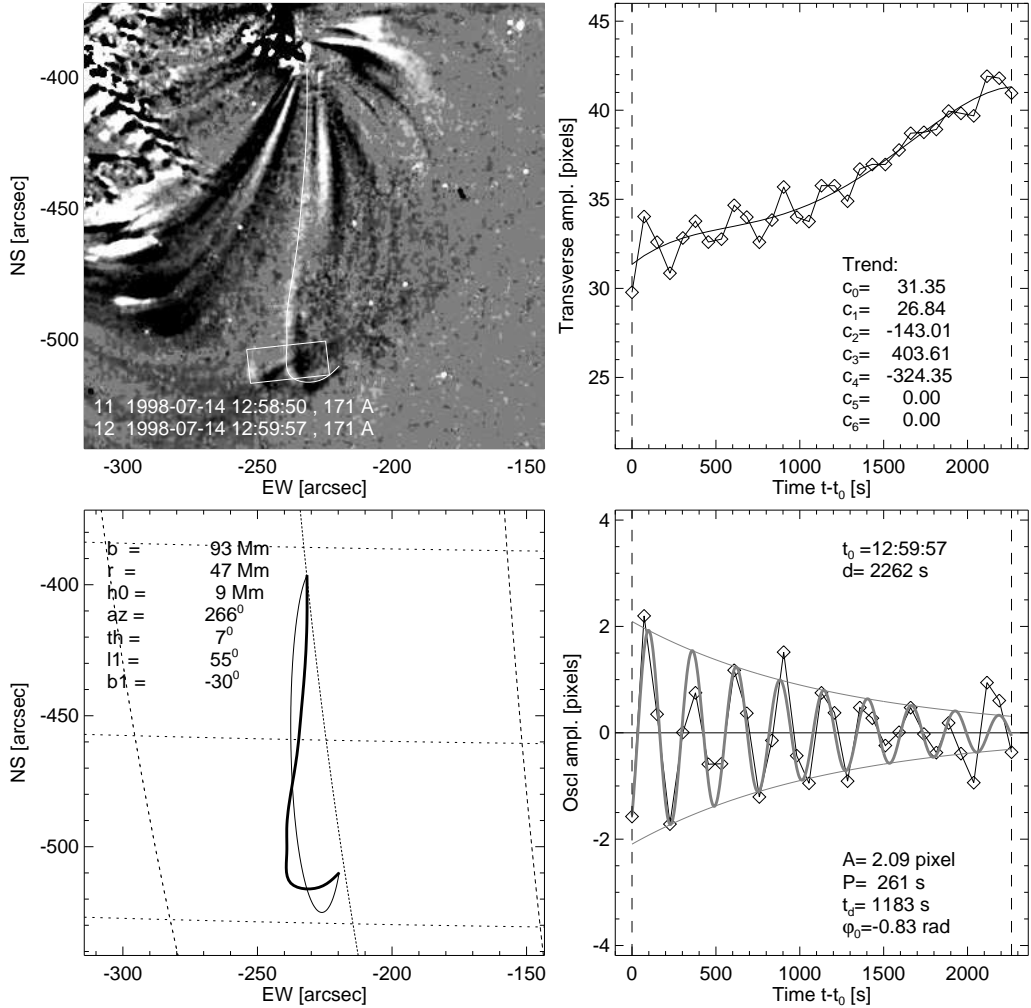}
\caption{Oscillation event No.~1a on 1998-Jul-14, 12:45 UT, analyzed in Aschwanden et al. (2002)
and in Nakariakov et al. (1999). A difference image is shown, where the transverse oscillation 
amplitude is analyzed in an area marked with a white box that is oriented perpendicular to the 
loop (top left panel). The 3D geometry of the loop is approximated with a coplanar circle
(thin line in bottom panel left). The oscillation amplitude is decomposed into a non-oscillatory
trend (top right panel) and into an oscillatory damped function (bottom right panel). For further
details see Aschwanden et al. (2002).}
\end{figure}

\begin{figure} 

\epsscale{.60}
\plotone{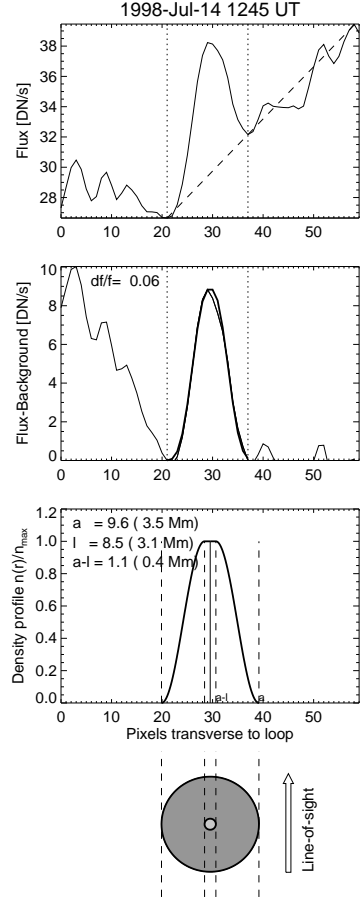}
\caption{The radial flux profile $F(r)$ for event 1a (see Fig.1) is shown (top panel), 
as function of the cross-sectional radius $r$ perpendicular to the loop and averaged 
along the loop within the white box shown in Fig.1 (top left panel). A linear background
to the oscillating loop is evaluated (dashed line in top panel) and subtracted (middle
panel). A trapezoidal density function with sinusoidal boundaries (Eq.1) with outer
radius $a$ and inner radius $a-l$ is shown (third panel) and fitted to the 
background-subtracted flux cross-section (middle panel) with proper line-of-sight integration
across the 2D density distribution (bottom).}
\end{figure}

\begin{figure} 
\plotone{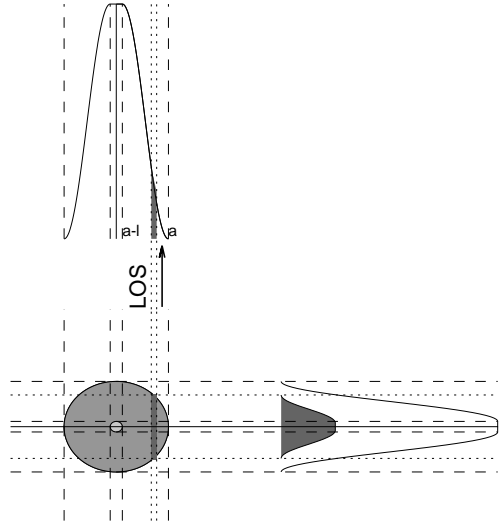}
\caption{The geometry of line-of-sight (LOS) integration is shown: The 2D density distribution
of a loop cross-section is contoured in grey-scale in the bottom left part, cross-sections
along the line-of-sight are projected to the right side, and the radial cross-section 
in transverse direction (in the plane of the sky) is projected to the top side.}
\end{figure}

\begin{figure} 
\plotone{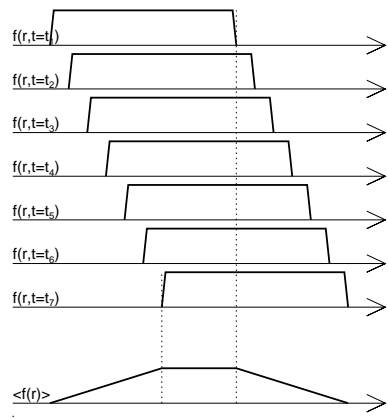}
\caption{The effect of time smearing is demonstrated for a near-rectangular cross-section
profile $F(t,t=t_i)$, $i=1,...,7$, which in superposition adds up to a more trapezoidal
cross-section profile $<F(r)>$.}
\end{figure}

\begin{figure} 
\plotone{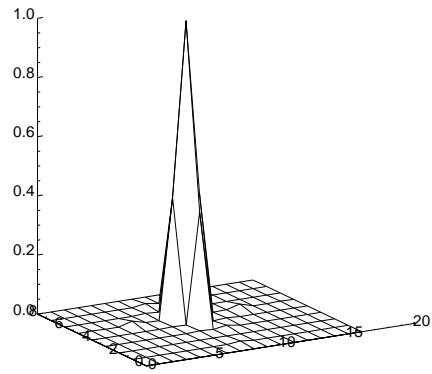}
\caption{The point-spread function of the TRACE 171 \ang\ filter measured by a 
{\sl Blind Iterative Deconvolution (BID)} algorithm. The mesh corresponds to
CCD pixels with a pixel size of 0.5" (courtesy of Richard Nightingale). }
\end{figure}

\begin{figure} 
\plotone{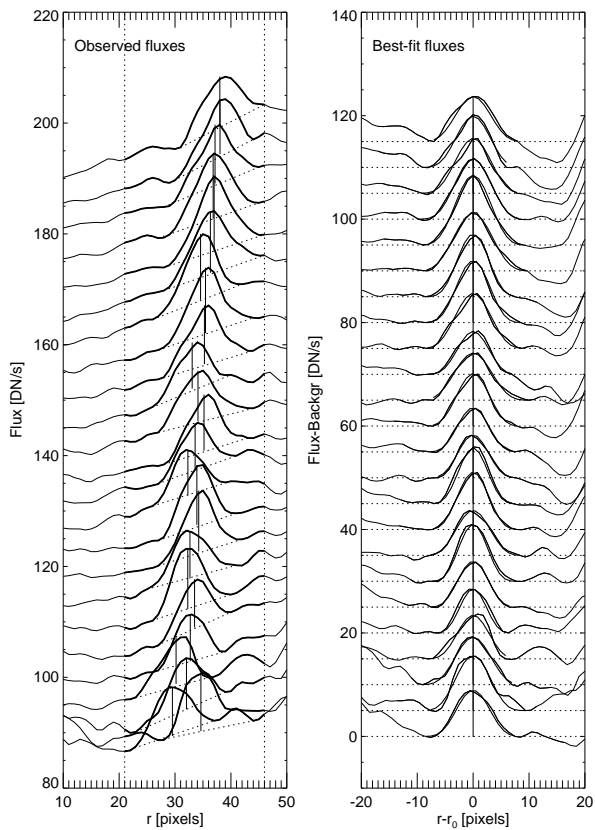}
\caption{Cross-sectional fluxes $F(r,t_i)$ for event 1a) are shown for time steps i=1,...,24 
(left panel), 
averaged within the white box area indicated in Fig.1 (top left). The flux profiles
are incrementally shifted by 5 flux units for clarity. The background fluxes are
indicated with dotted lines, the loop center location $r_0$ with a vertical solid line.
The best-fit flux profiles $F(r-r_0,t_i)$ are shown in the right panel (thick lines), based on the
density profile $n_e(r)$ (Eq.~1) obtained by forward-fitting of the line-of-sight
integration and instrumental response function. Note that the cross-sectional flux profile
is almost invariant during this entire time interval of the observed observation.}
\end{figure}

\begin{figure} 
\plotone{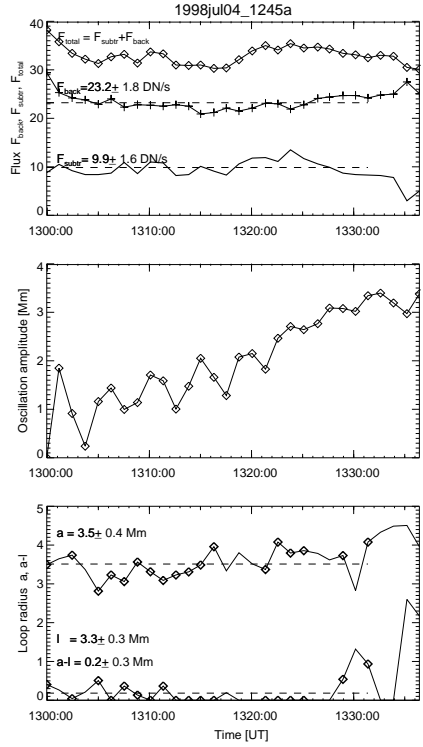}
\caption{Physical parameters of the best-fit cross-section profiles as function of time
for event 1a): Total flux $F_{total}(t)=F_{back}(t)+F_{loop}(t)$, background flux
$F_{back}(t)$, and loop flux $F_{loop}(t)$ (top panel); oscillation amplitude $A(t)$
(middle panel); and density profile parameters $a(t)$ and $l(t)$ (Eq.1) (bottom panel).
Acceptable fits (with deviations of $df/f<10\%$ in the flux profile) are marked with
diamonds.}
\end{figure}

\begin{figure} 
\plotone{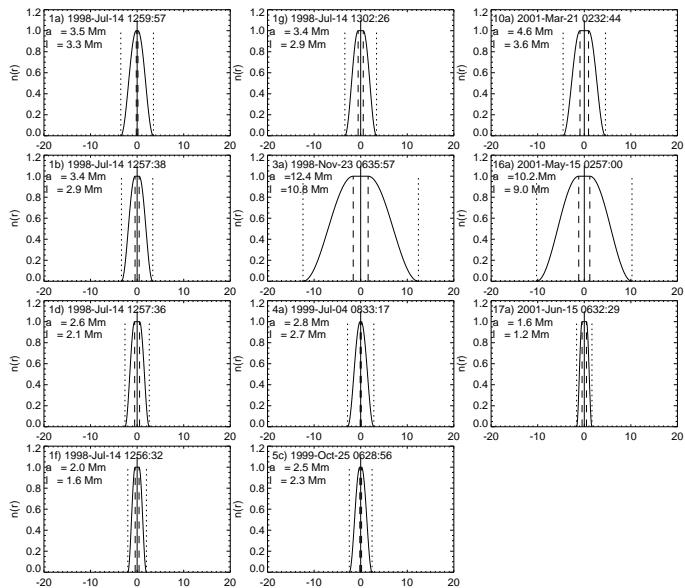}
\caption{Best-fit density profiles $n_e(r)$ as function of radial distance from loop
center $r$ [Mm] for all 11 events. The dates of the events and the best-fit parameters
$a$ and $l$ are indicated in each panel. Each profile is averaged from 10-30 timesteps
during the time interval of the observed loop oscillation.}
\end{figure}

\begin{figure} 
\plotone{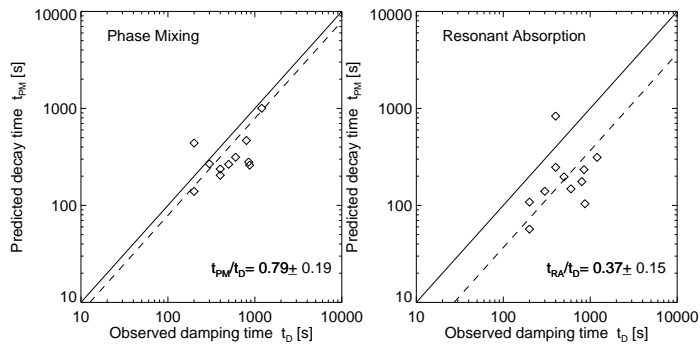}
\caption{{\sl Left panel:} Scaling of loop oscillation damping time $t_D$ with $(a\cdot P)$,
which does not show a close correlation. For damping by phase mixing a scaling of
$t_D \propto (P*l)^{2/3}$ is expected. {\sl Right panel:} Scaling of loop oscillation 
damping time $t_D$ with $(R/l)P$, for which a linear correlation is expected in the
framework of resonant absorption. The linear regression fit shows a slope of
$0.84\pm0.34$, which is consistent with the expected proportionality.}
\end{figure}

\begin{figure} 
\plotone{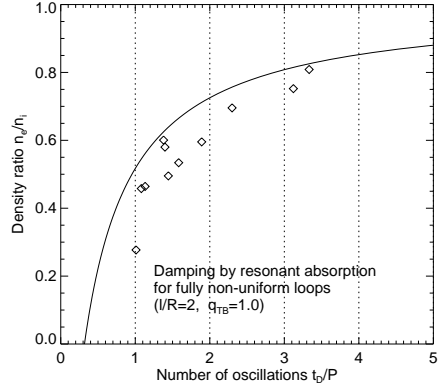}
\caption{The damping mechanism of resonant absorption provides a density
diagnostic of the density ratio $n_e/n_i$ (of external to internal density
in the oscillating loop) as function of the number of oscillation periods 
(measured by the ratio of damping time to the period, $N_{osc}=t_D/P$.
The plot shows the prediction in the fully non-uniform limit
($l/R=2$ and $q_{TB}=1.0$).}
\end{figure}

\end{document}